\documentclass[aps,prd,amsmath,twocolumn,showpacs]{revtex4}

\usepackage{epsfig}
\usepackage{graphics}
\usepackage{latexsym}
\usepackage{amsmath}
\usepackage{amssymb}
\usepackage{rotating}
\usepackage{subfigure}
\usepackage{bm}
\usepackage{color}
\usepackage{blindtext}
\begin{document}

\title{Study of isolated prompt photon production in $ p $-Pb collisions for the ALICE kinematics}

\author{Muhammad Goharipour}
\email{m.goharipour@semnan.ac.ir}

\author{Hossein Mehraban}
\email{hmehraban@semnan.ac.ir}

\affiliation{Faculty of Physics, Semnan University, Semnan P.O. Box 35131-19111, Semnan, Iran}

\date{\today}

\begin{abstract}

Prompt photon production is known as a powerful tool for testing perturbative QCD predictions
and also the validity of parton densities in the nucleon and nuclei especially of the gluon.
In this work, we have performed a detailed study on this subject focusing on the isolated
prompt photon production in $ p $-Pb collisions at forward rapidity at the LHC. The impact of
input nuclear modifications obtained from different global analyses by various groups
on several quantities has been investigated to estimate the order of magnitude of the
difference between their predictions. We have also studied in detail the theoretical uncertainties in
the results due to various sources. We found that there is a remarkable difference between the 
predictions from the nCTEQ15 and other groups in all ranges of photon transverse momentum $ p_\textrm{T}^\gamma $.
Their differences become more explicit in the calculation of the nuclear modification ratio 
and also the yield asymmetry between the forward and
backward rapidities rather than single differential cross sections. 
We emphasize that future measurements with ALICE will be very
useful not only for decreasing the uncertainty of the gluon nuclear modification, but also to accurately determine
its central values, especially in the shadowing region.

\end{abstract}


\maketitle

\section{Introduction}\label{sec:one} 
It is well known that the large momentum transfer processes play an important role 
in testing the perturbative quantum chromodynamics (pQCD) because 
the asymptotic freedom property allows us to apply the perturbative
techniques to make predictions for processes that are dominated by short-distance interactions.
In hadronic collisions, since photons couple in a pointlike fashion to the quark constituents of the colliding hadrons,
they provide an excellent probe for such purposes~\cite{Aurenche:1983ws,Owens:1986mp,Huston:1995vb,Aurenche:1998gv,Aurenche:2006vj}.
The study of prompt photons (photons not originating from meson decays) 
is also very useful to obtain direct information on the
parton distribution functions (PDFs), especially for the gluon~\cite{Aurenche:1988vi,Vogelsang:1995bg,Ichou:2010wc,dEnterria:2012kvo,Aleedaneshvar:2016dzb}.
Moreover, it has been established that prompt photon production in association with a
heavy quark (either charm or bottom) can be used for searching the intrinsic heavy quark 
components of the nucleon~\cite{Bednyakov:2013zta,Rostami:2016dqi}.

In heavy-ion collisions, the production of photons is recognized as an important
tool~\cite{Alam:1999sc} to study the fundamental properties of deconfined, strongly interacting matter, namely,
the quark gluon plasma (QGP)~\cite{Shuryak:1980tp,dEnterria:2006mtd} created in these collisions. 
Actually, photons can provide information on the
whole time evolution and dynamics of the medium because they are not accompanied by 
any final-state interaction. For example, measuring their transverse momentum distribution
can be used to estimate the temperature of the system. It is worth noting that, in
nucleus-nucleus collisions, direct photons come from a variety of different sources and 
can be divided into two categories: thermal photons that have a thermal origin and prompt photons
coming from cold processes. Usually, prompt photons are considered as a background
to thermal photons when we are looking for signals of the QGP. Thermal photons dominate the direct photon
spectrum at low photon transverse momentum $ p_\textrm{T}^\gamma $, while prompt
photons are the dominant photon source at high $ p_\textrm{T}^\gamma $.
Although in heavy-ion collisions there is no straightforward way to distinguish between thermal
and prompt photons experimentally, the fact that different processes are
dominant at different $ p_\textrm{T}^\gamma $ can help us to unfold the different contributions 
to the total observed yields. 

The measurement of direct photons in heavy-ion collisions has been performed so far in many experiments~\cite{Aggarwal:2000th,Adler:2005ig,Adare:2008ab,Afanasiev:2012dg,Adare:2014fwh,STAR:2016use,Chatrchyan:2012vq,Wilde:2012wc,Adam:2015lda,Aad:2015lcb}. For gold-gold collisions, 
one can refer, for example, to the PHENIX Collaboration measurements 
at center-of-mass energy of $ \sqrt{s}=200 $ GeV where the photons
have the transverse momentum $ 1\lesssim p_\textrm{T}\lesssim 20 $ GeV~\cite{Adler:2005ig,Afanasiev:2012dg}. 
The ALICE Collaboration has reported the first measurement
of a low-$ p_\textrm{T} $ direct photon at the LHC from lead-lead collisions at
$ \sqrt{s}=2.76 $ TeV~\cite{Wilde:2012wc,Adam:2015lda}. Such measurements have also been done 
by PHENIX in Au-Au collisions at the RHIC~\cite{Adare:2008ab,Adare:2014fwh}.
The measurement of prompt photon production at the LHC has also been performed by the
ATLAS~\cite{Aad:2015lcb} and CMS~\cite{Chatrchyan:2012vq} Collaborations 
in lead-lead collisions at $ \sqrt{s}=2.76 $ TeV in the ranges $ 22< p_\textrm{T}<280 $ GeV 
and $ 20< p_\textrm{T}<80 $ GeV, respectively. Note that in these kinematic regions, photons are expected to 
be dominantly produced in hard partonic collisions. For the case of d-Au collisions, we only have the
PHENIX measurement at $ \sqrt{s}=200 $ GeV~\cite{Adare:2012vn}. Despite all of these
experimental efforts, there is still no measurement of direct photon production
in $ p $-A collisions, though the ALICE measurement in $ p $-Pb collisions will be reported in the near future~\cite{Peitzmann:2016gkt}.

In order to calculate cross sections in any nuclear collision, one needs to know
the structure of the colliding nuclei. This structure can be described by 
nuclear parton distribution functions (NPDFs), similar to the PDFs in
the case of hadron structure. The required nuclear parton densities can be
extracted using the nuclear experimental data and DGLAP evolution equations
within the collinear factorization~\cite{Collins:1989gx,Brock:1993sz}. 
However, NPDFs cannot be well determined using the available 
nuclear deep inelastic scattering (DIS) and Drell-Yan experimental data compared 
to the free nucleon PDFs. Consequently, the obtained NPDFs from 
different global analyses by various groups~\cite{Hirai:2007sx,Schienbein:2009kk,Eskola:2009uj,deFlorian:2011fp,Khanpour:2016pph,Kovarik:2015cma,Hirai:2016ykc,Eskola:2016oht,Wang:2016mzo} 
have some considerable differences, 
both in behavior and uncertainty. This can lead to different results for
predictions of physical observables that are sensitive to NPDFs.
Since in nuclear collisions, prompt photons are produced in hard scatterings of incoming partons,
they can provide some information on parton densities in nuclei especially for the gluon PDF~\cite{Arleo:2007js,BrennerMariotto:2008st,Zhou:2010zzm,Arleo:2011gc,Helenius:2014qla}.
In this work, we investigate the impact of various recent NPDFs and their uncertainties
on the theoretical predictions of isolated prompt photon production in $ p $-Pb collisions at the LHC.
More emphasis will be placed on the recent nCTEQ15 NPDFs.

The contents of the present paper are as follows. In Sec.~\ref{sec:two}, we
discuss the gluon density of the proton and its nuclear modification for the Pb nucleus
and compare the predictions of various phenomenological groups at
different factorization scales. In Sec.~\ref{sec:three}, we briefly
describe the physics of prompt photon production. Particular attention is paid to 
its main concepts such as involved leading order (LO) and next-to-leading order (NLO) subprocesses, 
direct and fragmentation components of the cross section and the definition of isolation
cut. The isolated prompt photon production in $ pp $ and $ p $-Pb collisions at forward rapidity at the LHC
is studied in Sec.~\ref{sec:four}. The differential cross sections are calculated 
as functions of $ p_\textrm{T}^\gamma $ using various modern PDF and NPDF sets
to estimate the order of magnitude of the difference between their predictions.
Sec.~\ref{sec:five} is devoted to studying the theoretical uncertainties in the 
differential cross section of isolated prompt photon production due to various sources.
In Sec.~\ref{sec:six}, we calculate and compare the nuclear modification ratio $ R^\gamma_{p\textrm{Pb}} $
and the yield asymmetry between the forward and backward rapidities $ Y_{p\textrm{Pb}}^\textrm{asym} $
using different nuclear modifications. Finally, we summarize our results and conclusions in
Sec.~\ref{sec:seven}.

%
\section{gluon density of the proton and its nuclear modifications}\label{sec:two}
The accurate determination of PDFs is crucial
for all calculations of high-energy processes with initial hadrons,
whether within the standard model (SM) or when exploring
new physics. It is well known that the PDFs are nonperturbative
objects, and one needs to extract them from the global fits to hard-scattering
data. The reason for this is that they cannot be determined from the first principles of QCD, although 
their scale dependence is determined by the perturbative
DGLAP evolution equations. Nowadays, global analyses of PDFs are performed using a large number of
available precise experimental data from the DIS, Drell-Yan,
and collider experiments, and our knowledge of the quark
and gluon substructure of the nucleon has been improved to a large extent.
Consequently, the extracted PDFs by different analyst groups~\cite{Alekhin:2013nda,Ball:2014uwa,Harland-Lang:2014zoa,Jimenez-Delgado:2014twa,Abramowicz:2015mha,Dulat:2015mca,Accardi:2016qay,Butterworth:2015oua} are satisfactorily accurate
and also in a good agreement with each other. However, there are still some variations
in both their central values and uncertainties especially in the case of
gluon and sea quarks. Therefore, the study of those observables that are sensitive enough to
the specific parton distributions and can distinguish between them is of utmost importance
both experimentally and theoretically.

\begin{figure}[t!]
\includegraphics[width=0.48\textwidth]{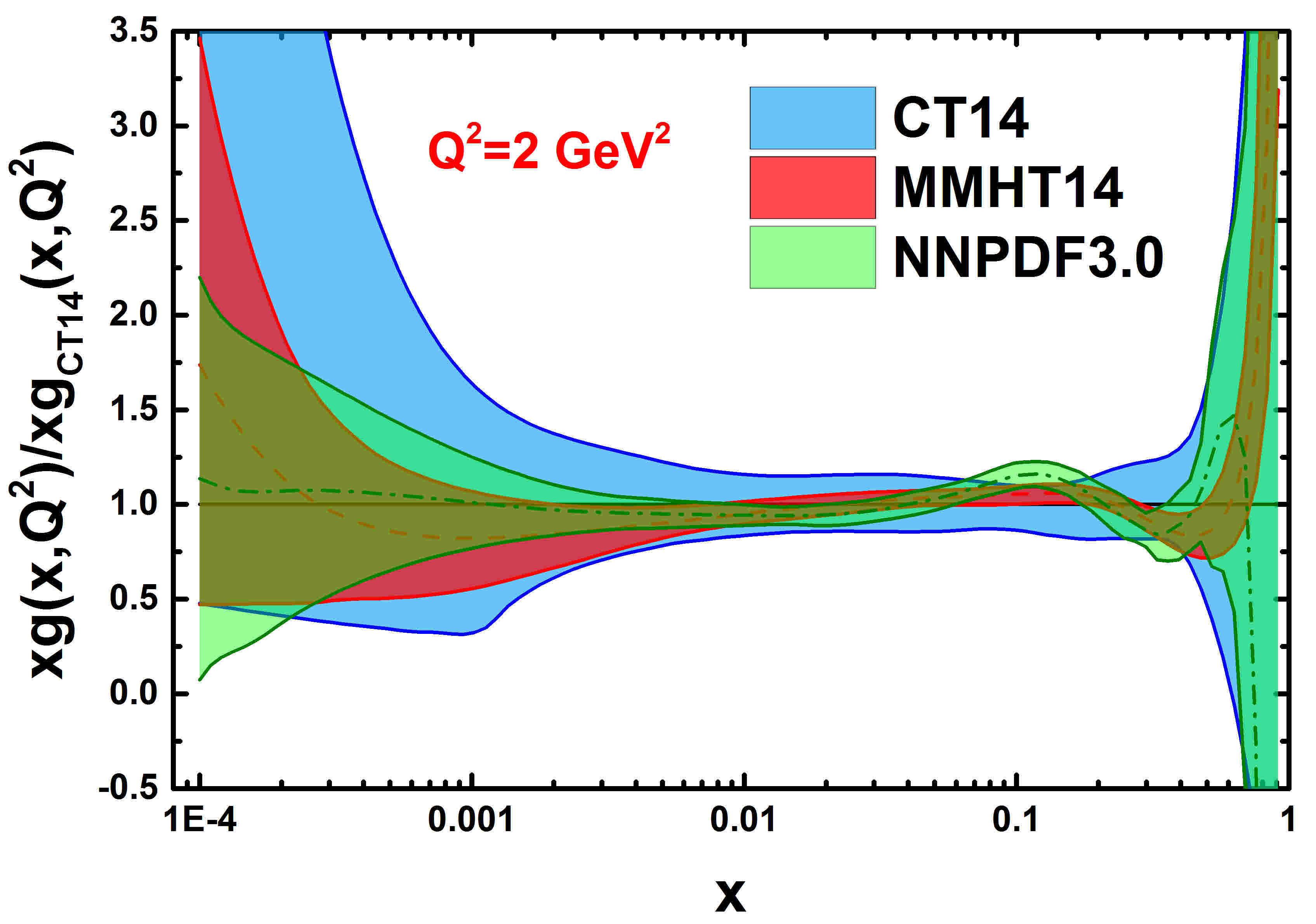}
\includegraphics[width=0.48\textwidth]{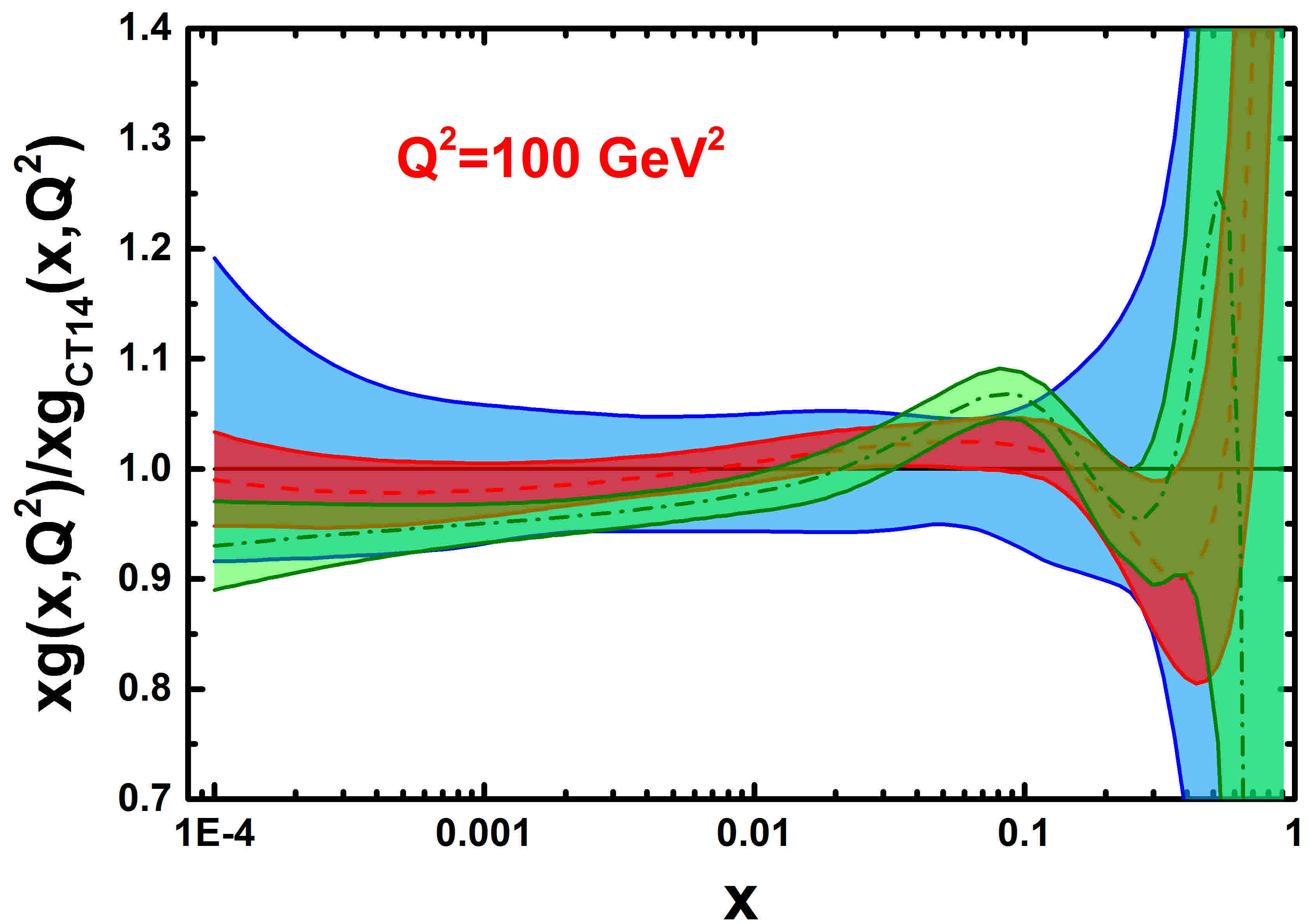}
\caption{ Ratio of the NLO gluon distributions with their uncertainties from 
various PDF sets: CT14~\cite{Dulat:2015mca} (blue band), MMHT14~\cite{Harland-Lang:2014zoa} (red band), 
and NNPDF3.0~\cite{Ball:2014uwa} (green band) to the CT14 central value at two
scales $ Q^2= $2 (top panel) and 100 (bottom panel) GeV$ ^2 $.  }
\label{fig:fig1}
\end{figure}
Since in this work we study the isolated direct photon production
at LHC energies, the gluon density is our favourite (see next section). Fig.~\ref{fig:fig1} shows a comparison
between the gluon distributions from the recent well-known phenomenological groups
namely, CT14~\cite{Dulat:2015mca}, MMHT14~\cite{Harland-Lang:2014zoa}, and NNPDF3.0~\cite{Ball:2014uwa}. 
We have plotted their NLO results in terms of the Bjorken scaling variable $ x $ at two
low (top panel) and high (bottom panel) scales $ Q^2= $2 and 100 GeV$ ^2 $.
The comparison has been made as the ratio to CT14 (as a reference PDF set) including PDF uncertainties
to make the differences between the results more clear.
As can be seen, the predictions of these groups for the gluon distribution differ significantly 
in some values of $ x $. Most differences occur in the small and large $ x $ regions.
Regardless of these regions, it may be of interest that the NNPDF3.0 has a significant enhancement at $ x $ of about 0.1.
Another conclusion that can be drawn from Fig.~\ref{fig:fig1} is that the gluon distribution from the CT14 
has a greater uncertainty (blue band)
than the MMHT14 (red band) and NNPDF3.0 (green band) in all values of $ x $.

In contrast to the PDFs, the results obtained for the NPDFs are not very satisfying due to the lack of experimental data. In fact,
since the current NPDF analyses~\cite{Hirai:2007sx,Eskola:2009uj,deFlorian:2011fp,Khanpour:2016pph,Schienbein:2009kk,Kovarik:2015cma,Hirai:2016ykc,Eskola:2016oht,Wang:2016mzo} are mainly constrained by DIS and Drell-Yan data,
only the quark nuclear modifications at fairly large values of $ x $ can be controlled as well.
Although the nuclear gluon distributions can be constrained indirectly via DGLAP evolution 
at higher orders of perturbation theory, we know that it is not enough for making accurate theoretical
predictions of physical observables that are sensitive to the gluon density. In this way, some 
phenomenological groups have used the inclusive pion production data from d-Au collisions at RHIC
in addition to the DIS and Drell-Yan data.
Very recently, Eskola {\it et al.}~\cite{Eskola:2016oht} have also used the LHC proton-lead
data in their analysis. However, due to the limited kinematic reach of data, the gluon modifications
and also their uncertainties obtained by various groups are very different 
in almost all values of $ x $. It should be noted that, at the moment, there is only one analysis
including the LHC $ p $-Pb data.

There are different approaches for determining the bound-proton
PDFs. Usually, they are defined in terms of nuclear modifications $ R_i^A $. To be more precise,
$ R_i^A $ are the scale-dependent ratios between the PDF of a proton inside a nucleus, $ f_i^{p/A} $,
and that in the free proton, $ f_i^p $,
\begin{equation}
R_i^A(x,Q^2)\equiv\frac{ f_i^{p/A}(x,Q^2)}{ f_i^p(x,Q^2)}.
\label{eq1}
\end{equation}
This approach has been used in the EPS09~\cite{Eskola:2009uj}, DSSZ~\cite{deFlorian:2011fp}, 
HKN07~\cite{Hirai:2007sx}, KA15~\cite{Khanpour:2016pph} and EPPS16~\cite{Eskola:2016oht} analyses. 
Note that, in this method, the extracted nuclear modifications
are dependent on the chosen PDFs of the free proton. For example, the HKN07 and EPS09
NPDFs are based on the MRST1998 free-proton~\cite{Martin:1998sq} and CTEQ6.1M~\cite{Stump:2003yu} sets, respectively.
However, there is another approach used by the nCTEQ group~\cite{Schienbein:2009kk,Kovarik:2015cma} in which the NPDF parametrisations
do not rely on a factorisation into a nuclear modification factor and free-proton PDFs.
Actually, in this approach, the NPDFs are parametrised directly as a function of $ x $
at the starting scale $ Q_0^2 $, and then an explicit $ A $ dependence is
introduced in the coefficients of their functional form.
It should be mentioned here that to obtain bound-neutron PDFs, one must assume the isospin symmetry. 
So, for the average up ($ u $) and down ($ d $) quark PDFs in a nucleus $ A $ with $ Z $ protons, we have
\begin{align}
u_A(x,Q^2)=\frac{Z}{A}R_u^A f_u^p+ \frac{A-Z}{A}R_d^A f_d^p\nonumber\\
d_A(x,Q^2)=\frac{Z}{A}R_d^A f_d^p+ \frac{A-Z}{A}R_u^A f_u^p.
\label{eq2}
\end{align}
\begin{figure}[t!]
\includegraphics[width=0.48\textwidth]{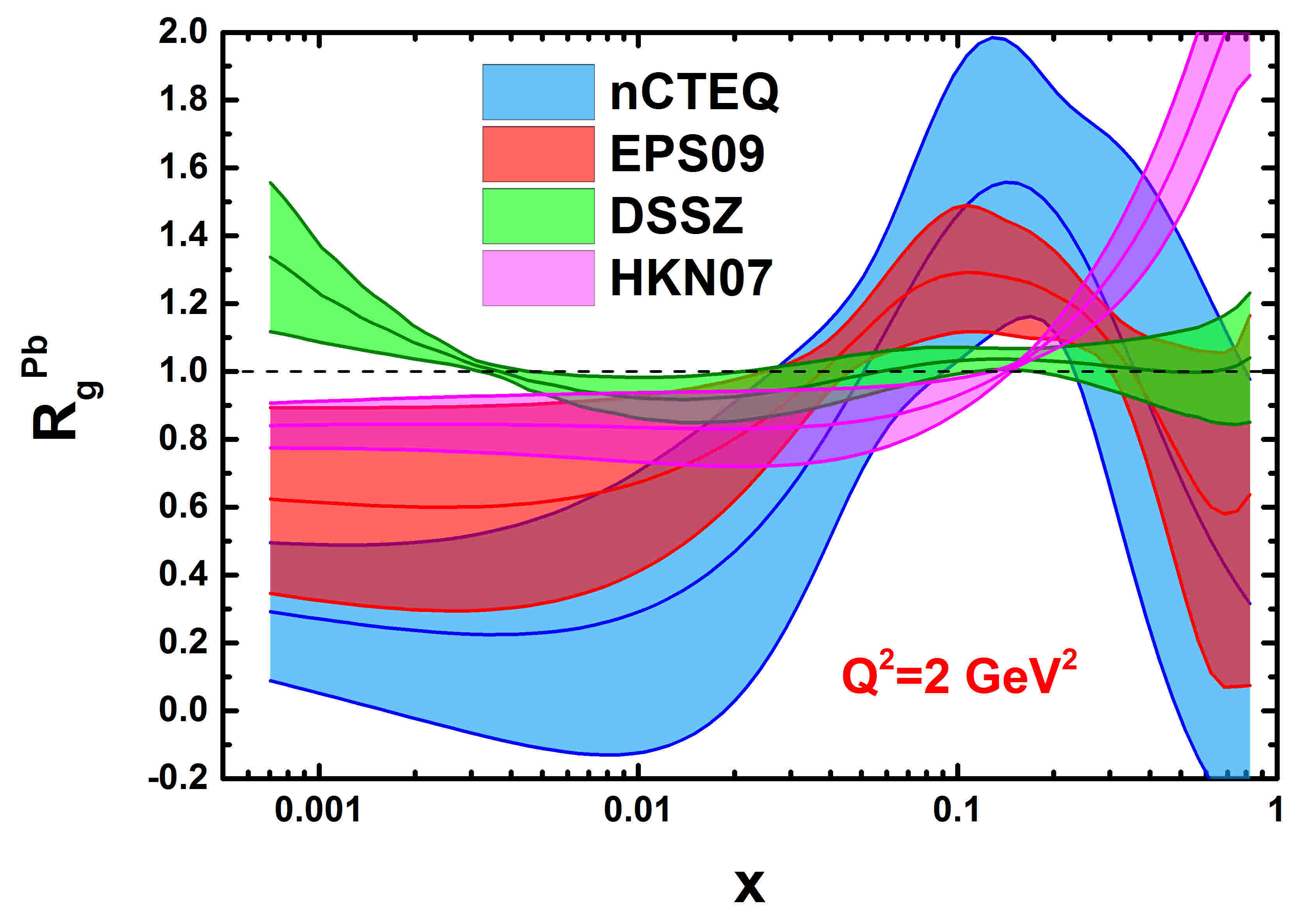}
\includegraphics[width=0.48\textwidth]{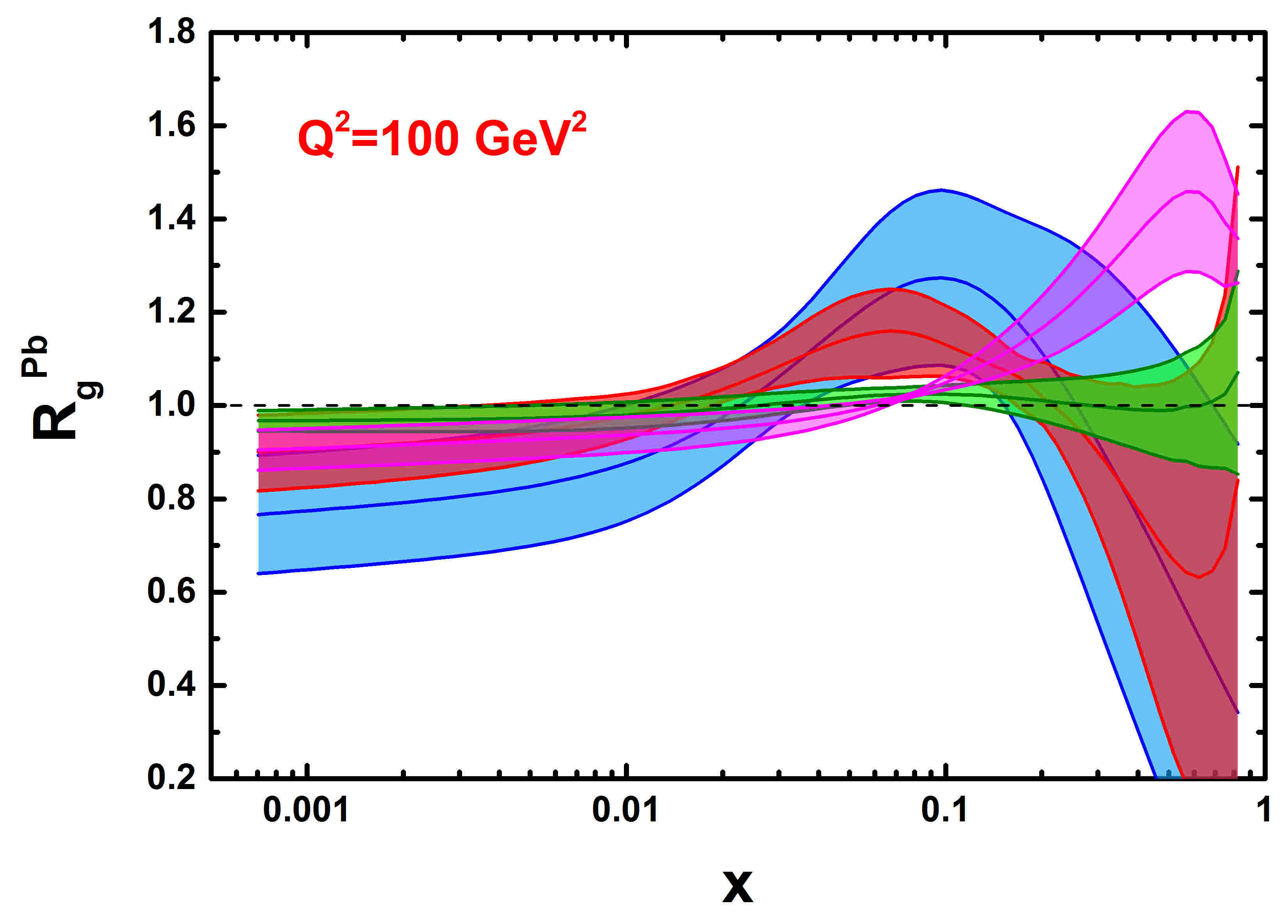}
\caption{ A comparison
between the nuclear modifications of the gluon PDF in a Pb-nucleus with uncertainties from 
the nCTEQ15~\cite{Kovarik:2015cma} (blue band), EPS09~\cite{Eskola:2009uj} (red band), 
DSSZ~\cite{deFlorian:2011fp} (green band) and HKN07~\cite{Hirai:2007sx} (pink band)
at $ Q^2= $2 (top panel) and 100 (bottom panel) GeV$ ^2 $.}
\label{fig:fig2}
\end{figure}
The nuclear modifications Eq.~\ref{eq1} can be divided into four areas
as a function of $ x $. A suppression for $ x\lesssim 0.01$ that is commonly referred to as shadowing.
Aanti-shadowing area in which $ R_i^A $ has an enhancement around $ x\sim0.1 $.
The EMC effect, a depletion at $ 0.3\lesssim x\lesssim0.7 $; and finally Fermi motion region in which
again $ R_i^A $ undergoes an excess towards $ x\rightarrow 1 $.
Fig.~\ref{fig:fig2} shows a comparison
between the nuclear modifications of the gluon PDF in a Pb-nucleus with their uncertainties from 
the nCTEQ15~\cite{Kovarik:2015cma} (blue band), EPS09~\cite{Eskola:2009uj} (red band), 
DSSZ~\cite{deFlorian:2011fp} (green band) and HKN07~\cite{Hirai:2007sx} (pink band)
at $ Q^2= $2 (top panel) and 100 (bottom panel) GeV$ ^2 $.
One can clearly see that there are remarkable differences between their central
values and uncertainties almost in whole ranges of $ x $. Among them,
the nCTEQ15 shows stronger shadowing, antishadowing and EMC effect.
Moreover, its prediction has wider error band than other groups in all values of $ x $.
The EPS09 has a similar treatment to nCTEQ15 but somewhat milder and also
with less uncertainty. Although the DSSZ dose not show the shadowing in
small values of $ x $ at $ Q^2= $2 GeV$ ^2 $, it appears in higher values of $ Q^2 $
due to the evolution effects. Nevertheless, the gluon shadowing in DSSZ is very small just like the
antishadowing, EMC effect and Fermi motion, so the DSSZ prediction for the
gluon nuclear modification of the lead nucleus stays around 1 in all regions.
Another interesting point that can be gained from Fig.~\ref{fig:fig2} is that the HKN07 does not show
the EMC effect neither for low nor high $ Q^2 $ values. In conclusion, one
can expect that these differences lead to the different results for predictions of physical observables that are sensitive to the gluon density
and so the gluon nuclear modification. Note that for the LHC with high values of center-of-mass energy $ \sqrt s $, 
depending on the transverse momentum $ p_\textrm{T} $ and pseudorapidity $ \eta $ of photons, 
various $ x $ regions can be explored in the target and the projectile. 
Actually, with the naive LO $ 2\rightarrow2 $ kinematics,
the momentum fractions typically probed by direct photon production
are
\begin{equation}
x_{1,2}\approx \frac{2p_\textrm{T}}{\sqrt s} e^{\pm \eta}.
\label{eq3}
\end{equation}
Therefore, for a given $ \sqrt s $ and $ p_\textrm{T} $, and for the case in which $ \eta $ increases, the process
becomes sensitive to parton densities at smaller $ x_2 $ (target) and larger $ x_1 $ (projectile).  
In the next section, we briefly review the
physics of the prompt photon production in the LHC collisions as an excellent
probe for the gluon distribution of the proton and its corresponding nuclear modification.

%
\section{physics of prompt photon production}\label{sec:three}
For more than three decades, many studies have been done on the
prompt photon production
~\cite{Aurenche:1983ws,Aurenche:1987fs,Owens:1986mp,Huston:1995vb,Aurenche:1989gv,Baer:1990ra,Berger:1990et,Gordon:1994ut,Aurenche:1998gv,Catani:2002ny,Belghobsi:2009hx,Aurenche:2006vj,Schwartz:2016olw,Lipatov:2016wgr,Helenius:2013bya,Fontannaz:2001ek,Odaka:2015uqa,Cleymans:1994nr, Skoro:1999rg, Bolzoni:2005xn, Lipatov:2005wk,Jezo:2016ypn,Campbell:2016lzl,Siegert:2016bre,Benic:2016uku,Kohara:2015nda}. 
In this section, we briefly discuss the prompt photon physics and related topics.
By definition, ``prompt photons" are those photons that arise from processes during the collision
and are not produced from the decay of hadrons, such as $ \pi_0 $, $ \eta $, etc. produced
at large transverse momenta. Forward prompt photons consist of two types of photons: 
direct and  fragmentation photons. Direct photons that behave as high-$ p_\textrm{T} $
colorless partons are produced predominantly from the
initial hard scattering processes of the colliding quarks or
gluons. Fragmentation photons behave as a kind of 
hadron, i.e. they are produced as bremsstrahlung emitted by a
scattered parton, from the fragmentation of high-$ p_\textrm{T} $ quarks and gluons
which are produced in primary hard partonic collisions
or from the interaction of a scattered parton with the
medium created in heavy-ion collisions~\cite{Turbide:2007mi,Vitev:2008vk}.
Although the direct and fragmentation components of the prompt photon 
cross section described above cannot be measured separately in the experiments, the theoretical
calculations can be performed completely separately. In this way, the cross section for
the inclusive prompt photon production in a collision of
hadrons $ h_1 $ and $ h_2 $ can be written generally as follows
\begin{equation}
d\sigma_{h_1h_2}^{\gamma +X}= d\sigma_{h_1h_2}^{\textrm{D~} \gamma +X}+d\sigma_{h_1h_2}^{\textrm{F~} \gamma +X},
\label{eq4}
\end{equation}
where D and F refer to the direct and fragmentation parts, respectively,
and $ X $ indicates the inclusive nature of the cross section. On the other hand, using
collinear factorization~\cite{Collins:1989gx,Brock:1993sz},
the cross section for the inclusive production of a hard elementary particle $ k $ 
in $ h_1h_2 $ collisions can be calculated as
\begin{align}
d\sigma_{h_1h_2}^{k +X}= \sum_{i,j} f_{i}^{h_1}(x_1,M^2)&\otimes
f_{j}^{h_2}(x_2,M^2)\nonumber\\ 
& \otimes d\hat{\sigma}_{i,j}^{k +X^\prime}(\mu^2,M^2,M_F^2),
\label{eq5}
\end{align}
where $ f_{i}^{h_1}(x_1,M^2) $ and $ f_{j}^{h_2}(x_2,M^2) $ are 
the PDFs of parton species $ i $ and $ j $ inside the projectile ($ h_1 $) and target ($ h_ 2 $),
respectively, at momentum fractions $ x_1 $ and $ x_2 $ and factorization scale $ M $ of the initial-state parton distributions.
In the above equation, $ \otimes $ is a convolution integral over $ x_1 $ and $ x_2 $. Moreover, the
partonic pieces $ d\hat{\sigma}_{i,j}^{k +X^\prime} $ denoted by $ M $, 
renormalization scale $ \mu $ and factorization scale $ M_F $ of the photon fragmentation function can be calculated as a perturbative
expansion in the strong ($ \alpha_s $) and electroweak ($ \alpha $) couplings (Note that 
the partonic pieces $ d\hat{\sigma}_{i,j}^{k +X^\prime} $ do not have any dependence on
the factorization scale $ M_F $ in the Born approximation. However, 
such dependence appears in the calculation of the higher-order corrections
to both the direct and fragmentation components from which final-state collinear singularities 
have been subtracted according to the $ \overline{\textrm{MS}} $ factorization scheme~\cite{Catani:2002ny}.) 
As usual, $ X^\prime $ indicates that in the calculation of $ d\hat{\sigma}_{i,j}^{k +X^\prime} $, 
we must integrate over everything else but the photon. Then, we can calculate the direct component of
the prompt photon production cross section in Eq.~\eqref{eq4} using Eq.~\eqref{eq5} and
assuming particle $ k $ as a photon $ \gamma $.

The experimentally measured prompt photons also include 
the fragmentation photons emitted through collinear fragmentation of a 
parton that is itself produced with a large transverse momentum. The fragmentation
component of the prompt photon production cross section in Eq.~\eqref{eq1} 
can be calculated as follows
\begin{align}
d\sigma_{h_1h_2}^{\textrm{F~}\gamma +X}=& \sum_{i,j,k} f_{i}^{h_1}(x_1,M^2)\otimes
f_{j}^{h_2}(x_2,M^2)\nonumber\\ 
& \otimes d\hat{\sigma}_{i,j}^{k +X^\prime}(\mu^2,M^2,M_F^2)\otimes D_{\gamma/k}(z,M_F^2).
\label{eq6}
\end{align}
In this equation, $ D_{\gamma/k}(z,M_F^2) $ is the parton-to-photon fragmentation function (FF)
where $ z $ is the fractional momentum over which the last convolution is taken. Actually,
in the calculation of the fragmentation contribution, occurrence of some singularities 
including a final-state quark-photon collinear singularity or final-state multiple collinear
singularities at higher orders is inevitable. These singularities are resummed and absorbed
into FFs of the photons. In this case, since the fragmentation functions behave roughly
as $ \alpha/\alpha_s (M_F^2)$~\cite{Aurenche:1983ws,Owens:1986mp}, the perturbatively calculable pieces
related to the partonic subprocesses, $ d\hat{\sigma}_{i,j}^{k +X^\prime} $,
can be of the order of $ \alpha_s^2 $ for LO and $ \alpha_s^3 $ for NLO
parton production. Consequently, the fragmentation contributions of the cross section
remain of the same order as the direct contributions.

Now we are in a position to introduce all partonic subprocesses that contribute to the 
prompt photon production cross section at
LO and NLO approximation. At LO, there are two Born-level subprocesses: the Compton scattering
$ q(\bar q)g \rightarrow \gamma q(\bar q) $ and annihilation $ q\bar q \rightarrow \gamma g $.
The importance of these subprocesses clearly depends on the type of collisions. 
Actually, in $ pp $ collisions at RHIC and LHC,
the $ q\bar q $ annihilation channel has a small contribution to the cross sections in all kinematic regions,
whereas at the Tevatron this channel is also considerable.
At NLO, there are more contributing subprocesses including $ q(\bar q)g \rightarrow \gamma gq(\bar q) $, 
$ q\bar q \rightarrow \gamma gg $, and other subprocesses 
from the virtual corrections to the Born-level processes. It is worth pointing out in this context that since at the LHC
the $ q\bar q $ annihilation is suppressed compared to other subprocesses, and on the
other hand, the gluon distribution is dominant rather than the sea quark
distributions at small $ x $, the prompt photon production provides direct information on
the proton gluon distribution. It should also be taken into account that because of the high
center-of-mass energy, the photon production at the
LHC probes values of $ x $ that are considerably smaller than at the Tevatron.

In order to reject the background of photons coming
from the decays of hadrons such as $ \pi_0 $, $ \eta $ produced in the collision
that are not considered as prompt photons by definition, an isolation criterion
is required. Various isolation criteria have been used so far in related 
studies~\cite{Catani:2002ny,Kunszt:1992np,Glover:1993xc,Frixione:1998jh}.
The most common criterion which can be also
implementable at the partonic level is the cone criterion~\cite{Catani:2002ny}.
According to the cone isolation criterion, a photon is considered as an isolated photon
if, in a cone of radius $ R $ in rapidity $ y $ and azimuthal angle $ \phi $ around
the photon direction,
\begin{equation}
(y-y_\gamma)^2+(\phi-\phi_\gamma)^2\leq R^2,
\label{eq7}
\end{equation}
the amount of accompanying hadronic transverse energy $ E_\textrm{T}^{\textrm{had}} $
is smaller than some finite value  $ E_{\textrm{T}\textrm{~max}} $,
\begin{equation}
E_\textrm{T}^{\textrm{had}}\leq E_{\textrm{T}\textrm{~max}.}
\label{eq8}
\end{equation}
Both $ R $ and $ E_{\textrm{T}\textrm{~max}} $ are chosen by the experiment and
$ E_{\textrm{T}\textrm{~max}} $ is presented as a fixed value or a fixed fraction
of the transverse momentum of the photon $ p_\textrm{T}^\gamma $ or, more generally, as a function of $ p_\textrm{T}^\gamma $.

One of the main differences between the direct and fragmentation photons is that 
a direct photon will most probably be separated from the hadronic environment,
whereas a fragmentation photon, except for the case in which the photon
carries away most of the momentum of the fragmenting parton, is most probably accompanied by hadrons.
In this way, since the isolation cut discards the prompt photon events that have too much
hadronic activity around the photon, and on the other hand, the fragmentation photons
are emitted collinearly to the parent parton, it is expected that the 
isolation cut reduces the fragmentation component. Now, considering what was said above,
in the next section, we calculate and study in detail the prompt photon production in $ pp $ and $ p $-Pb collisions 
at center-of-mass energy of $ 8.8 $ TeV and for the forward rapidities corresponding to 
the ALICE kinematics~\cite{Peitzmann:2016gkt}.

%
%
\section{Study of isolated prompt photon production at ALICE}\label{sec:four}
The LHC allows us to investigate the behaviour of SM particles in a qualitatively new
energy region via measuring the production of various particles such as the
$ W $ and $ Z $ bosons in association
with jets~\cite{Aad:2014rta} or a heavy flavor quark~\cite{Aad:2011kp,Aad:2013ysa} 
and also the isolated prompt photon whether inclusively~\cite{Khachatryan:2010fm,Aad:2010sp,Aad:2013zba,Aad:2016xcr,Aaboud:2017cbm} 
or in association with jets~\cite{ATLAS:2012ar,Aad:2013gaa}. In the previous section, we presented  
main topics related to the prompt photon production in hadron collisions.
In this section, we present theoretical predictions for 
the isolated prompt photon production in $ pp $ and $ p $-Pb collisions 
at $ \sqrt{s}=8.8 $ TeV corresponding to 
the ALICE kinematics~\cite{Peitzmann:2016gkt}. All calculations are performed 
here, and subsequent sections are based on the J\textsc{et}P\textsc{hox} Monte Carlo program~\cite{Catani:2002ny,Aurenche:2006vj,Belghobsi:2009hx} 
which includes both direct and fragmentation processes 
and also allows us to study the isolation cut. We include all diagrams up to 
LO and NLO of QED and QCD coupling, respectively,
defined in the $ \overline{\textrm{MS}} $ renormalization scheme. Within the J\textsc{et}P\textsc{hox} framework, 
it is also possible to compute direct and fragmentation 
parts as distinct, however, the NLO calculations are performed at the
parton level and do not account for hadronization effects.
It should be noted that, in our numerical calculations
performed in this section and also Sec.~\ref{sec:six}, we use
set II of the NLO Bourhis-Fontannaz-Guillet (BFG) FFs of photons~\cite{Bourhis:1997yu} 
for calculating the fragmentation component of the cross sections [Eq.~\eqref{eq6}]. 
Moreover, the renormalization ($ \mu $), factorization ($ M $) and fragmentation
($ M_F $) scales are set to the photon transverse momentum ($ \mu=M=M_F=p_\textrm{T}^\gamma $).
The FFs and scale uncertainties are studied separately in the next section in addition to the theoretical uncertainties
due to NPDFs. As a last point, note that the fine-structure constant ($ \alpha _{\textrm{EM}}$) 
is set to the J\textsc{et}P\textsc{hox} default of 1/137.

\begin{figure}[t!]
\centering
\includegraphics[width=0.48\textwidth]{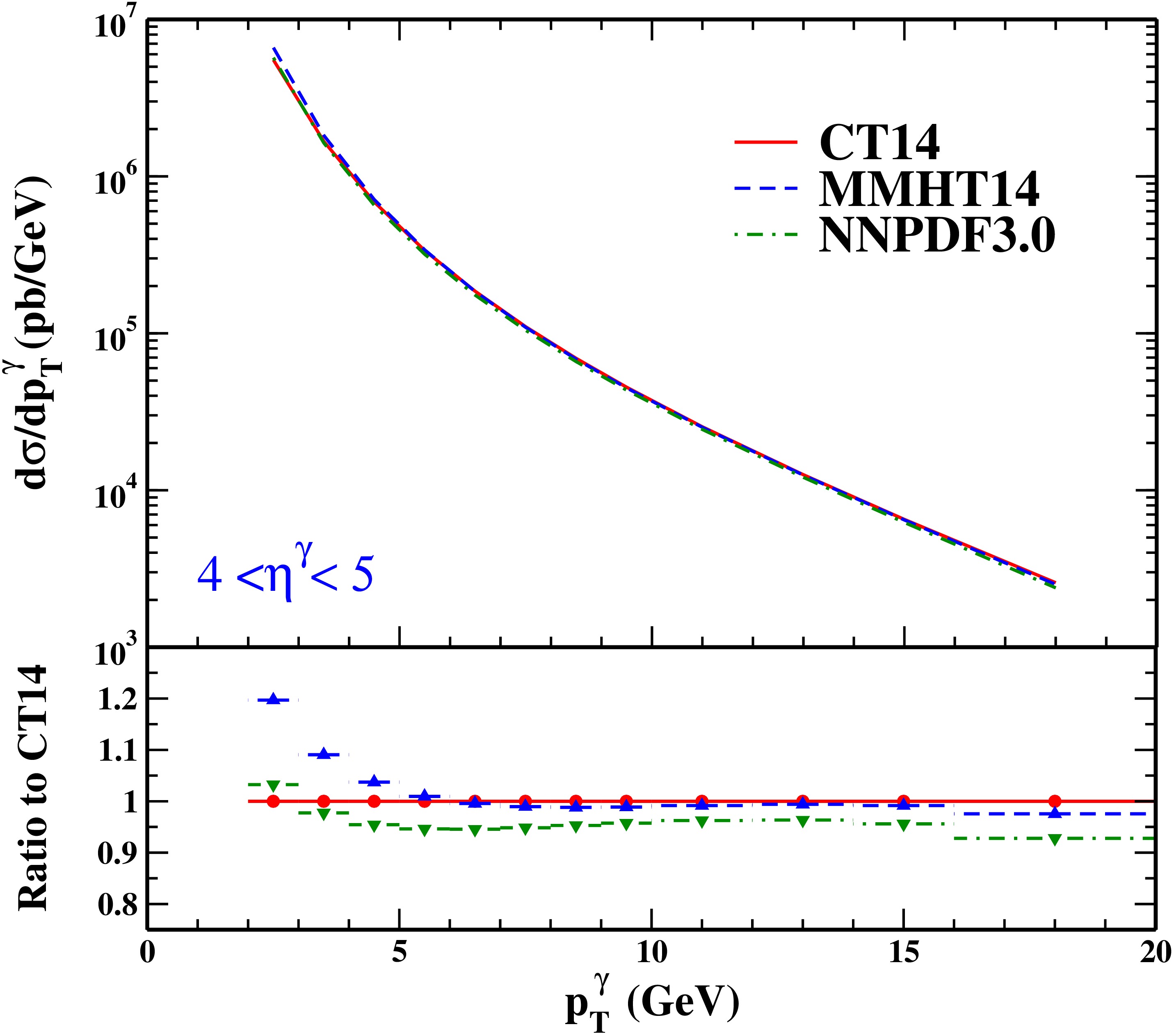}
\caption{A comparison of the NLO theoretical predictions for the
differential cross section of isolated prompt photon production in $ pp $ collisions as a function of $ p_\textrm{T}^\gamma $ 
using three various NLO PDFs of CT14~\cite{Dulat:2015mca} (red solid),
MMHT14~\cite{Harland-Lang:2014zoa} (blue dashed) and 
NNPDF3.0~\cite{Ball:2014uwa} (green dotted-dashed) at $ \sqrt{s}=8.8 $ TeV
for $ 4<\eta^\gamma<5 $. The ratios of the results to the CT14 prediction have been shown in the bottom panel.}
\label{fig:fig3}
\end{figure}

Now we are in a position to calculate the isolated prompt photon production in
$ pp $ collisions at $ \sqrt{s}=8.8 $ TeV theoretically, using various modern 
PDF sets CT14~\cite{Dulat:2015mca}, MMHT14~\cite{Harland-Lang:2014zoa} and NNPDF3.0~\cite{Ball:2014uwa}
introduced in Sec.~\ref{sec:two}.
In this way, we can estimate the variation of the results due to the different PDF sets.
Note that for each group, the NLO PDF sets with $ \alpha_s(M_Z)=0.118 $ are taken
by virtue of the LHAPDF package~\cite{Whalley:2005nh}. Fig.~\ref{fig:fig3} shows the obtained differential cross sections
as a function of $ p_\textrm{T}^\gamma $ in the kinematic range $ 2<p_\textrm{T}^\gamma<20 $ GeV for 
the forward region $ 4<\eta^\gamma<5 $. We should note that in calculating the cross sections, we have used a tighter isolation
cut, $ E_\textrm{T}^{\textrm{had}}< 2 $, with $ R=0.4 $ [see Eqs.~\eqref{eq7} and~\eqref{eq8}].
As can be seen, all predictions are in good agreement with each
other almost in all regions of $ p_\textrm{T}^\gamma $. However, in order to investigate in more detail
the differences between the predictions in various regions of $ p_\textrm{T}^\gamma $, 
we have plotted their ratios to the CT14 prediction in the bottom panel of Fig.~\ref{fig:fig3}.
The only significant difference occurs at $ p_\textrm{T}\simeq 2 $, where the MMHT14 prediction
differs from the CT14 and NNPDF3.0 ones up to 20\%. Besides this, we can
state that the differences between them are less than 10\% in all values of $ p_\textrm{T}^\gamma $.

\begin{figure}[!]
\centering
\includegraphics[width=8.6cm]{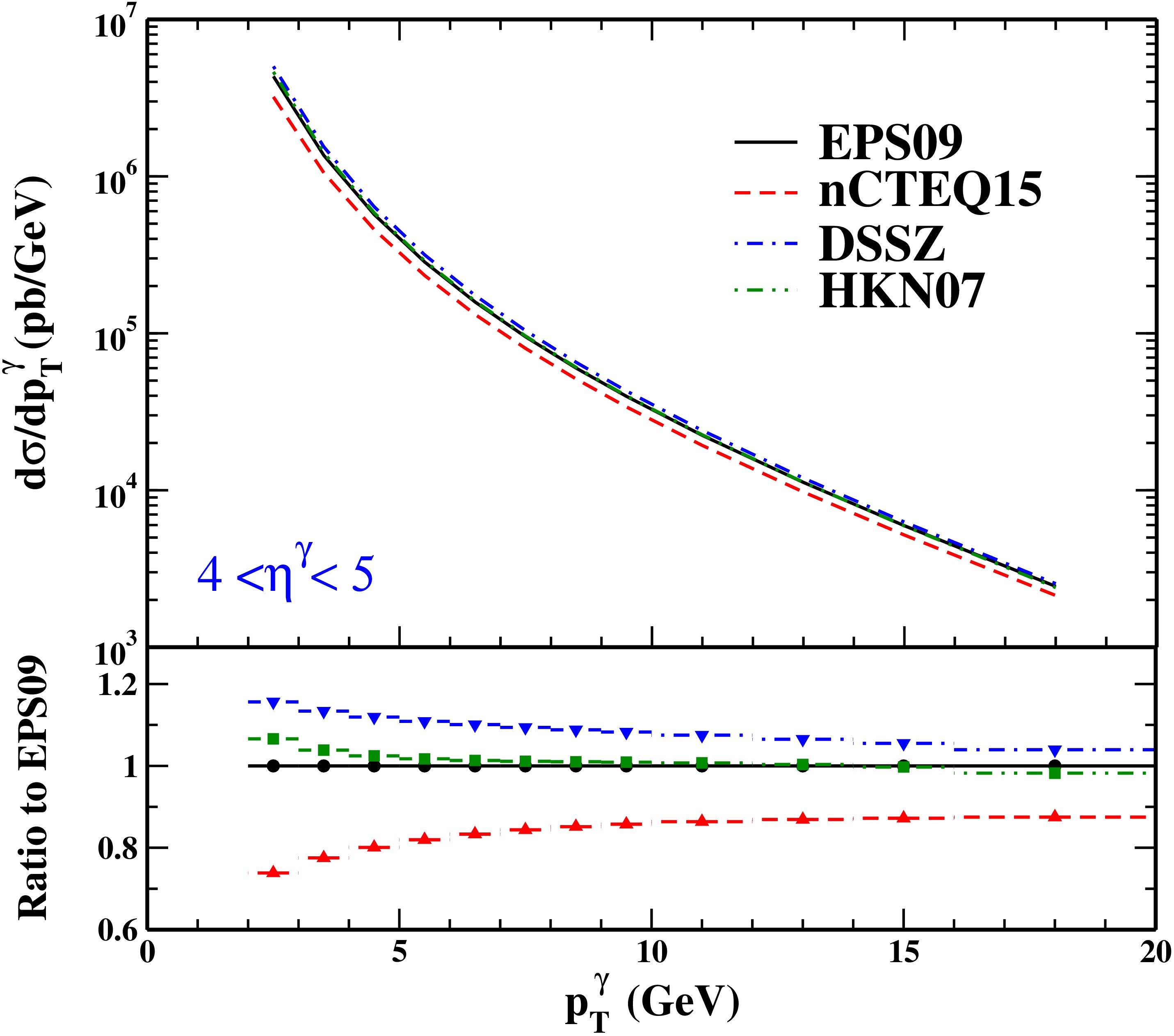}
\caption{A comparison of the NLO theoretical predictions for the
differential cross section of isolated prompt photon production in $ p $-Pb collisions as a function of $ p_\textrm{T}^\gamma $
using the EPS09~\cite{Eskola:2009uj} (black solid), nCTEQ15~\cite{Kovarik:2015cma} (red dashed), 
DSSZ~\cite{deFlorian:2011fp} (blue dotted-dashed) and HKN07~\cite{Hirai:2007sx} (green dotted-dotted-dashed) 
nuclear modifications and the CT14 free-proton PDFs~\cite{Dulat:2015mca} at $ \sqrt{s}=8.8 $ TeV
for $ 4<\eta^\gamma<5 $. The ratios of the results to the EPS09 prediction are shown in the bottom panel.}
\label{fig:fig4}
\end{figure}
In Sec.~\ref{sec:two}, we have shown that the nuclear modifications of the gluon distribution from various
phenomenological groups differ from each other, to a large extent, almost in all values of $ x $.
Now, as a next step, we calculate the NLO differential cross section of the isolated prompt photon
production in $ p $-Pb collisions at $ \sqrt{s}=8.8 $ TeV in order to study the impact of input nuclear modifications on the
final results and estimate the order of magnitude of the difference between their predictions. 
To this end, we take the nuclear modifications, Eq.~\eqref{eq1}, from 
the nCTEQ15~\cite{Kovarik:2015cma}, EPS09~\cite{Eskola:2009uj}, 
DSSZ~\cite{deFlorian:2011fp} and HKN07~\cite{Hirai:2007sx} and choose the CT14 PDF sets for
the free-proton PDFs. The calculations are performed again for 
the forward region $ 4<\eta^\gamma<5 $. The obtained results have been compared in Fig.~\ref{fig:fig4} 
as a function of $ p_\textrm{T}^\gamma $ in the kinematic range of $ 2<p_\textrm{T}^\gamma<20 $ GeV.
In the bottom panel, we have shown their ratios to the EPS09 prediction.
As a result, one can clearly see that these groups have different predictions for 
isolated prompt photon production at the ALICE kinematics. Although the HKN07 is
in a good agreement with EPS09 in all values of $ p_\textrm{T}^\gamma $, the DSSZ and nCTEQ15
have significant deviations especially at smaller values of $ p_\textrm{T}^\gamma $. Overall,
the nCTEQ15 which presented the newest modern NPDFs (among the sets considered in this work) 
has the greatest difference from the others and its prediction is placed below them.
Note that according to Eq.~\eqref{eq3}, for $ p $-Pb collisions and kinematics used here, the photon
probes the NPDFs in small values of $ x_2 $ corresponding to the shadowing region
in Fig.~\ref{fig:fig2}. Then, due to the large differences observed in Fig.~\ref{fig:fig4},
measurements of isolated prompt photon production
at the ALICE kinematics can be really helpful in constraining the gluon nuclear modifications and determining
their best central values in the shadowing region.
The differences between the various gluon modifications from different groups can be made more explicit 
if one calculates the minimum bias nuclear modification ratio for $ p $-Pb collisions
and also the yield asymmetry between the forward and
backward rapidities. We study these quantities separately in Sec.~\ref{sec:six}.
In the next section, we investigate the theoretical uncertainties in the 
differential cross section of isolated prompt photon production due to NPDF, scale and FF uncertainties.

%
\section{Study of theoretical uncertainties}\label{sec:five} 
In the previous section we calculated the cross section of isolated prompt photon
production in $ p $-Pb collisions using various nuclear modifications of PDFs for the Pb nucleus.
Now, it is important and also interesting to calculate the theoretical uncertainties in
the results with respect to the various sources. The most important sources of uncertainties are the PDF, NPDF, scale and FF 
uncertainties. Since the theoretical uncertainties of the free-proton PDFs have been
studied before in many papers concerning the isolated prompt photon production in $ pp $ collisions,
and on the other hand, since we are interested here in $ p $-Pb collisions and thus the impact of NPDFs
on the cross section, we ignore the study of PDF uncertainties (note, however, that PDFs have
smaller uncertainties than NPDFs). 

To study the NPDF uncertainties, we choose the nuclear modifications from the nCTEQ15~\cite{Kovarik:2015cma}
as our baseline which has greatest uncertainties in comparison with other groups
according to Fig.~\ref{fig:fig2}. The theoretical uncertainties of 
nuclear modifications can be obtained as usual using the 32 error sets of the nCTEQ15 parametrisation
as shown in Fig.~\ref{fig:fig2} by the blue band.
For calculating such uncertainties to any physical quantity related to NPDFs 
such as the isolated prompt photon production considered here, one must vary the error sets 
and calculate the deviations from the central result (the best-fit value) and then the contribution
to the size of the upper and lower errors via
\begin{align}
\delta^+ X= \sqrt{\sum_i [\max (X^{(+)}_i-X_0,X^{(-)}_i-X_0,0)]^2},\nonumber\\
\delta^- X= \sqrt{\sum_i [\max (X_0-X^{(+)}_i,X_0-X^{(-)}_i,0)]^2}.
\label{eq9}
\end{align}

The other important sources of theatrical uncertainties in the cross section of isolated prompt photon
production are the uncertainties due to the scale variations. As mentioned, we have set the renormalization,
factorization and fragmentation scales as $ \mu=M=M_F=p_\textrm{T}^\gamma $ in all calculations performed
in the previous section. Although, no optimal scale choice is possible for the prediction of the inclusive 
photon cross section in the region of phase space of interest~\cite{Blair}, it is accepted that the predictions
can be reliably made by this choice. Moreover, we accept that the 
theoretical uncertainty due to scale variations can be calculated by changing these scales
by a factor 2 and comparing the $ 2p_\textrm{T}^\gamma $ and $ \frac{1}{2}p_\textrm{T}^\gamma $ results.
However, the more correct method in the calculation of scale uncertainties is
the combination of both incoherent and coherent scale variations.
In this method, an incoherent variation means varying the scales independently
by a factor of 2 around the central value so that one scale is varied while keeping the other two equal to $ p_\textrm{T}^\gamma $,
and a coherent variation means varying the scales simultaneously by a factor of 2 around the central value as before.
Finally, we can calculate the total scale uncertainty by adding in quadrature all obtained uncertainties.

Since, as mentioned in Sec.~\ref{sec:three}, prompt photon production consists of
both direct and fragmentation contributions, it is inevitably related to the FFs that appeared
in the fragmentation component [see Eq.~\eqref{eq6}]. Then, the part of the theoretical uncertainties
in its cross section comes from the FFs uncertainties. Unfortunately, at present,
because of the lack of experimental data for inclusive photon production in $ e^+e^- $ annihilation
as a best source to constrain the photon fragmentation functions, our knowledge about them
especially for the case of gluon fragmentation to photons is not satisfactory. Note that, for example, 
the ALEPH and HRS data on $ \rho $ production have been used in the BFG parametrisations~\cite{Bourhis:1997yu}.
Consequently, their gluon fragmentation to photons has large uncertainty and is parametrised with sets I and II.
Although the significant difference between these two sets appears at low scales and
may not matter at LHC energies, it is interesting to see to what extent the isolated prompt 
photon cross section is impacted by changing the FF set. In the previous section, the BFG set II
was used to calculate the fragmentation component of the cross sections. In this section, we use set I to
estimate the FF uncertainties by comparing the results with the previous ones.

\begin{figure}[t!]
\centering
\includegraphics[width=8.6cm]{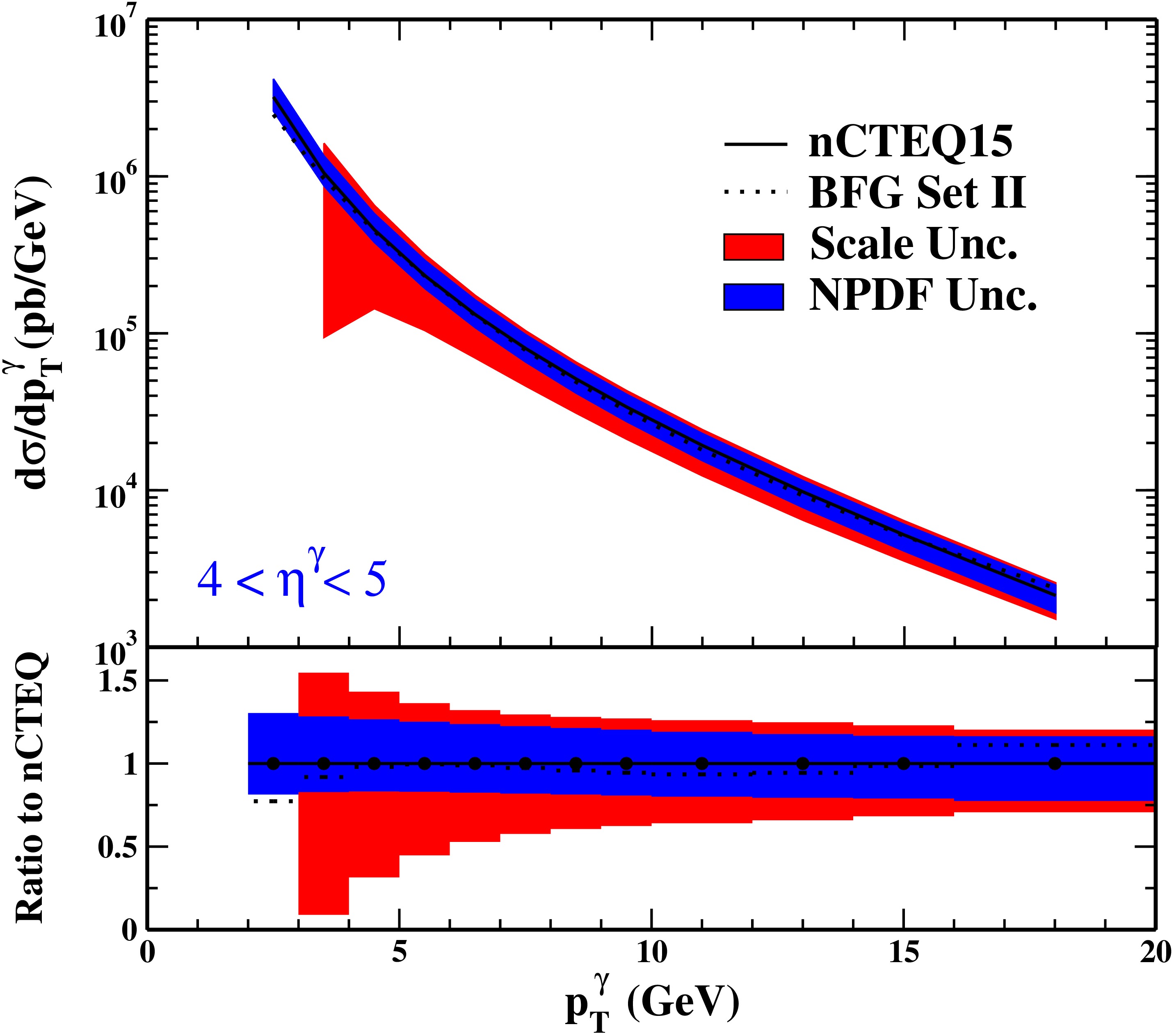}
\caption{A comparison between the  NPDF, scale and
FF uncertainties in the differential cross section of isolated prompt photon production in $ p $-Pb collisions at $ \sqrt{s}=8.8 $ TeV 
as a function of $ p_\textrm{T}^\gamma $ for the forward region $ 4<\eta^\gamma<5 $. 
The black solid and dotted curves are the nCTEQ15~\cite{Kovarik:2015cma} predictions using  
FFs from the BFG sets II and I~\cite{Bourhis:1997yu}, respectively.  The
red band represents the scale uncertainties. The nCTEQ15 NPDF uncertainties are shown
by the blue band. The bottom panel shows the ratios to the nCTEQ15 central prediction.}
\label{fig:fig5}
\end{figure}
Now, according to what was said above, we are ready to calculate the theoretical uncertainties
in the cross section of isolated prompt photon production in $ p $-Pb collisions at $ \sqrt{s}=8.8 $ TeV due to the NPDF, scale and
FF uncertainties. Fig.~\ref{fig:fig5} shows the results obtained as a function of $ p_\textrm{T}^\gamma $ 
for the forward region $ 4<\eta^\gamma<5 $
using the nCTEQ15~\cite{Kovarik:2015cma} parametrizations as inputs for the nuclear modifications.
In this figure, the dotted curve represents the results obtained using the BFG set I for the FFs and the
red and blue bands represent the scale and NPDF uncertainties, respectively. Note that the black solid curve
corresponds to the results obtained in the previous section using the nCTEQ15 central set and also the BFG set II.
As before, the ratios to the nCTEQ15 central prediction are shown in the bottom panel. As can be seen,
there is not significant difference between the predictions obtained using the FFs of BFG sets I and II.
Some deviations are seen just at low and large values of $ p_\textrm{T}^\gamma $. The scale uncertainties
are dominant rather than NPDF uncertainties in all ranges of $ p_\textrm{T}^\gamma $, and they become 
very large at low $ p_\textrm{T}^\gamma $. Note that we have plotted the scale uncertainties for $ p_\textrm{T}^\gamma> $3 GeV
because for its smaller values, the cross section becomes unphysical when one sets the scales to
the lower value $ \mu= \frac{1}{2}p_\textrm{T}^\gamma $. It is also worth noting here that if one considers only
the coherent scale variations to calculate the scale uncertainties, a narrow error band is obtained 
in almost all $ p_\textrm{T}^\gamma $ regions, so the scale uncertainties do not even exceed 5\%.

%
\section{Nuclear modification and forward-to-backward ratios}\label{sec:six}
\begin{figure}[!]
\centering
\includegraphics[width=8.6cm]{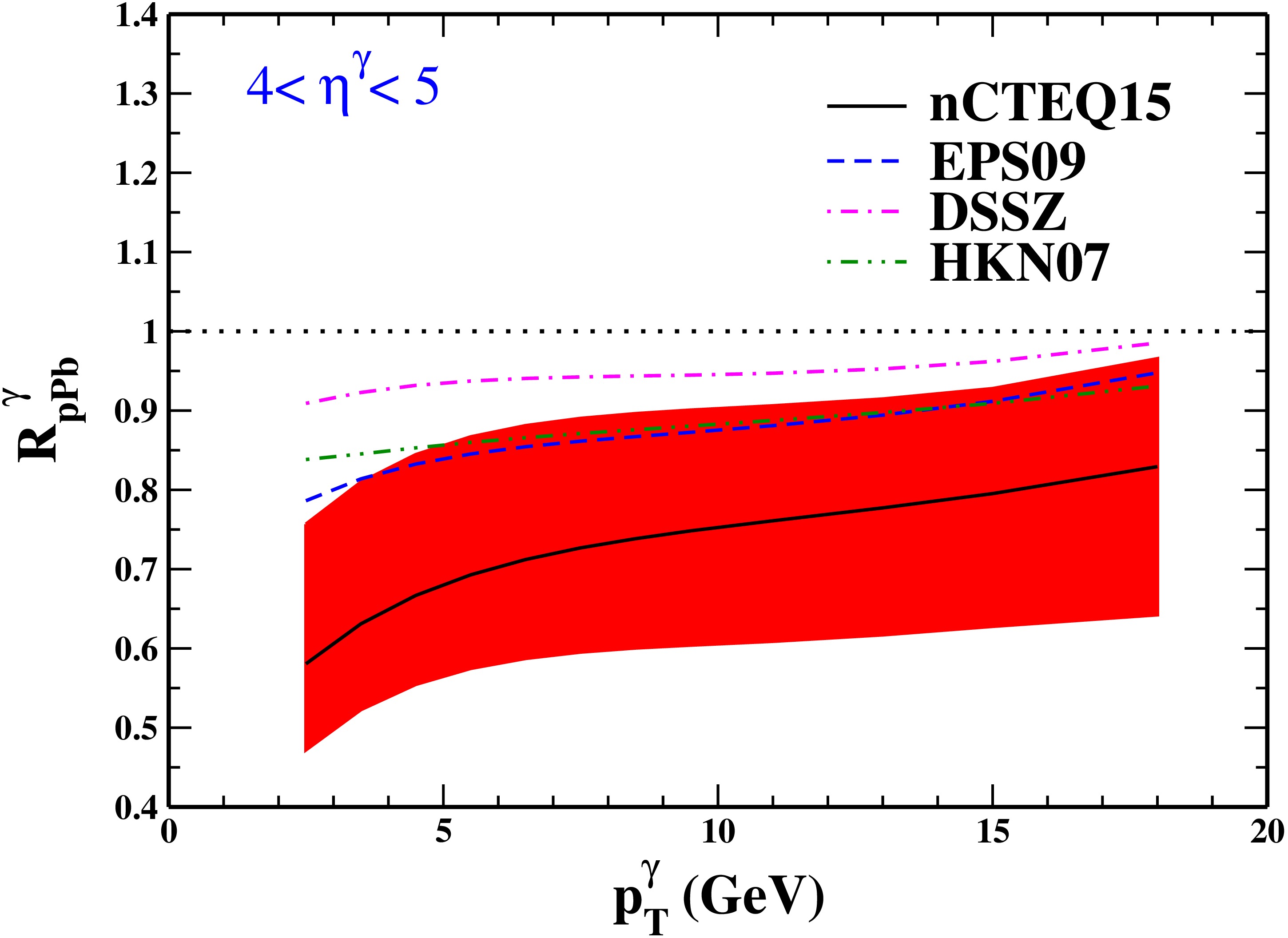}
\caption{A comparison between the nuclear modification ratios $ R_{p\textrm{Pb}}^\gamma $
for $ p $-Pb collisions at $ \sqrt{s}=8.8 $ TeV and $ 4<\eta^\gamma<5 $
using the nCTEQ15~\cite{Kovarik:2015cma} (black solid), EPS09~\cite{Eskola:2009uj} (blue dashed), 
DSSZ~\cite{deFlorian:2011fp} (pink dotted-dashed) and HKN07~\cite{Hirai:2007sx} (green dotted-dotted-dashed) 
nuclear modifications and the CT14 free-proton PDFs~\cite{Dulat:2015mca}.
The red band corresponds to the nCTEQ15 NPDF uncertainties.}
\label{fig:fig6}
\end{figure}

In the previous section, we found that the theoretical uncertainties due to the scale variations
can be very large especially at low values of $ p_\textrm{T}^\gamma $ if one uses the method 
in which the combination of both incoherent and coherent scale variations is considered.
On the other hand, there are also PDF and FF uncertainties besides the NPDF uncertainties.
In this way, it is very desirable to have a quantity that is not only more sensitive to the 
nuclear modifications, but one also in which the other sources of theoretical uncertainties are canceled
to a large extent. In this regard, the minimum bias nuclear modification ratio is a good choice~\cite{Helenius:2014qla}.
For the prompt photon production in $ p $-Pb collisions at the LHC, it is defined as
\begin{equation}
R_{p\textrm{Pb}}^\gamma\equiv \frac{d\sigma/dp_\textrm{T} (p\textrm{+Pb}\rightarrow \gamma +\textrm{X})}{208\times d\sigma/dp_\textrm{T} (p+p\rightarrow \gamma +\textrm{X})}.
\label{eq10}
\end{equation}
Note that because of the isospin effect, the nuclear modification ratio Eq.~\eqref{eq10} is not normalized to 1
when no nuclear modifications in the parton densities are assumed. However, the isospin effect
becomes more important whenever the valence quark sector of the nuclei is probed.
The predictions for $ R_{p\textrm{Pb}}^\gamma $ at $ 4<\eta^\gamma<5 $ and 
$ \sqrt{s}=8.8 $ TeV are shown in Fig.~\ref{fig:fig6} in the kinematic range $ 2<p_\textrm{T}^\gamma<20 $ GeV.
In this figure, the central nCTEQ15~\cite{Kovarik:2015cma} prediction has been
shown as a black solid line and the red band corresponds to its uncertainty range,
while the EPS09~\cite{Eskola:2009uj}, DSSZ~\cite{deFlorian:2011fp} and HKN07~\cite{Hirai:2007sx} 
predictions are represented by the blue dashed, pink dotted-dashed and
green dotted-dotted-dashed curves, respectively. It should be noted that for free-proton PDFs,
whether in the numerator or denominator of Eq.~\eqref{eq10}, we have again used the CT14 PDFs~\cite{Dulat:2015mca}.
As can be seen, there is a remarkable difference between the nCTEQ15 prediction and other groups
in all the ranges of $ p_\textrm{T}^\gamma $. Note that the DSSZ prediction is not even within the large
error band of the nCTEQ15. It indicates that future measurements with ALICE will be very
useful not only for decreasing the uncertainty of the gluon nuclear modification, but also to accurately determine
its central values in the shadowing region.

Although the nuclear modification ratio Eq.~\eqref{eq10} is a quantity with a high sensitivity to 
the nuclear modifications in PDFs and is also largely indifferent to the
PDF, FF and scale uncertainties, we can define the other quantity that does not require
a $ p+p $ baseline measurement with the same $ \sqrt{s}$. It is the yield asymmetry between the forward and
backward rapidities, which for $ p $-Pb collisions is defined as follows:
\begin{equation}
Y_{p\textrm{Pb}}^\textrm{asym} \equiv \frac{d\sigma/dp_\textrm{T} (p\textrm{+Pb}\rightarrow \gamma +\textrm{X})\mid_{\eta\in[\eta_1,\eta_2]}}{d\sigma/dp_\textrm{T} (p\textrm{+Pb}\rightarrow \gamma +\textrm{X})\mid_{\eta\in[-\eta_2,-\eta_1]}}.
\label{eq11}
\end{equation}
Such an observable has the advantage
that it is free from the absolute normalization uncertainty included due to involving
the Glauber modeling~\cite{Miller:2007ri} for the cases in which the luminosity for the collected data sample is not measured. 
Also some correlated systematic uncertainties can be expected to cancel.

\begin{figure}[!]
\centering
\includegraphics[width=8.6cm]{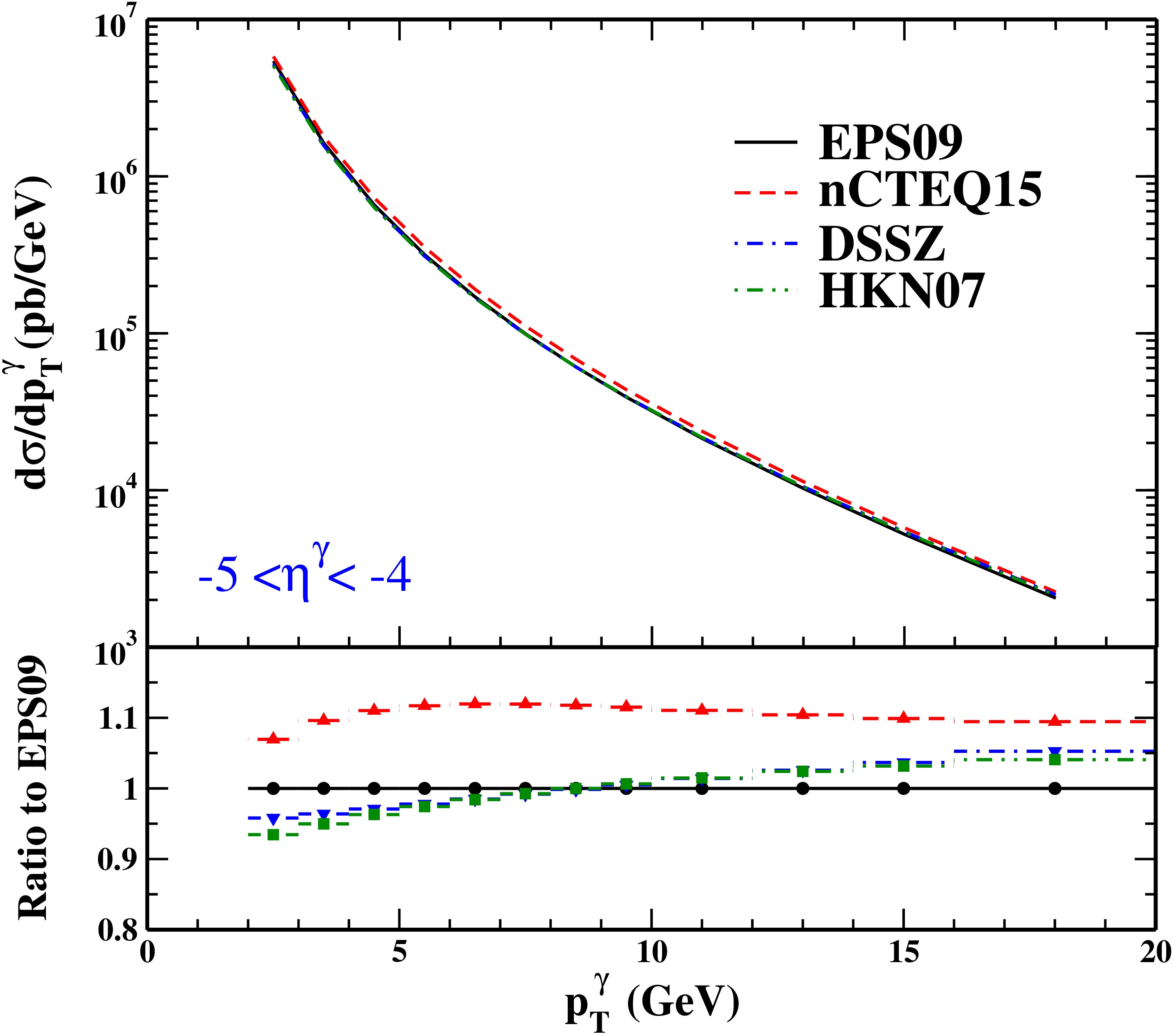}
\caption{Same as Fig.~\ref{fig:fig4}, but for the backward region $ -5<\eta^\gamma<-4 $. }
\label{fig:fig7}
\end{figure}
According to Eq.~\eqref{eq3}, the isolated photon production at backward rapidities will be sensitive to the
nuclear antishadowing and EMC effect of NPDFs. This means that in this kinematic region
the nuclear valence quark modifications also become important since the sensitivity of the cross section
to them is raised towards larger values of $ x_2 $. It is now interesting to calculate the
differential cross sections and also nuclear modification ratios at backward rapidities before calculating
the forward-to-backward ratios. Fig.~\ref{fig:fig7} shows the NLO theoretical predictions for the
differential cross section of isolated prompt photon production as a function of $ p_\textrm{T}^\gamma $
using the EPS09 (black solid), nCTEQ15 (red dashed), DSSZ (blue dotted-dashed) and HKN07 (green dotted-dotted-dashed) 
nuclear modifications and CT14 free-proton PDFs at $ \sqrt{s}=8.8 $ TeV
for the backward region $ -5<\eta^\gamma<-4 $. 
The ratios of the results to the EPS09 prediction have been shown in the bottom panel.
In analogy to the forward region $ 4<\eta^\gamma<5 $ (see Fig.~\ref{fig:fig4}), the differences between the predictions are
smaller; thus, the DSSZ and HKN07 have a very similar behavior in all values of $ p_\textrm{T}^\gamma $,
and their predictions only have a little difference from the EPS09 prediction at low and
high $ p_\textrm{T}^\gamma $ regions. Although the differences between the nCTEQ15
prediction and other groups are somewhat decreased in the backward region, there are
still considerable deviations. Note also that unlike the forward case, the nCTEQ15 prediction
is placed on the top of other predictions due to its larger nuclear antishadowing,
and it becomes closer to them towards the EMC effect region.

\begin{figure}[!]
\centering
\includegraphics[width=8.6cm]{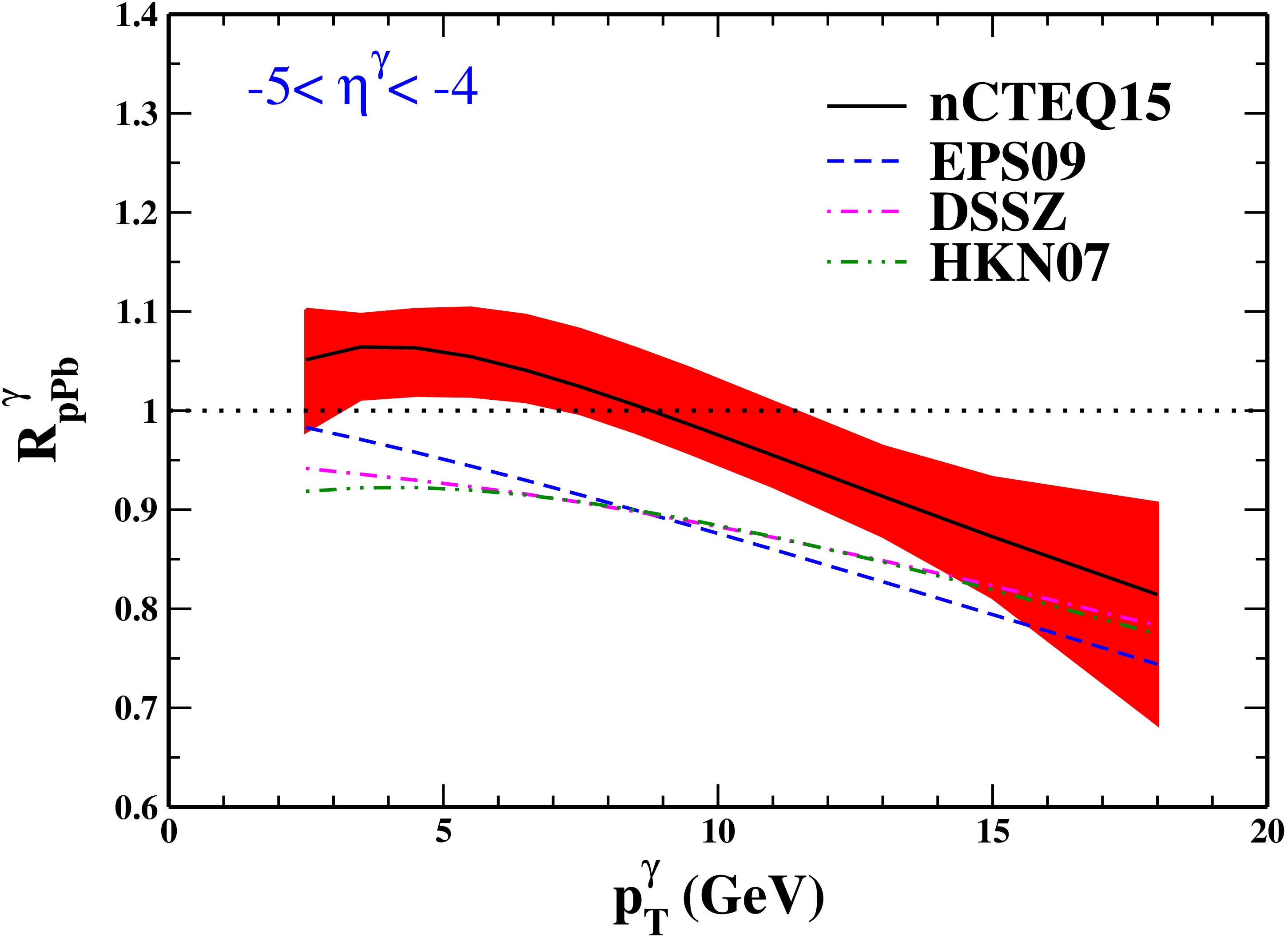}
\caption{Same as Fig.~\ref{fig:fig6}, but for the backward region $ -5<\eta^\gamma<-4 $. }
\label{fig:fig8}
\end{figure}

The nuclear modification ratios $ R_{p\textrm{Pb}}^\gamma $ for $ p $-Pb collisions 
at $ \sqrt{s}=8.8 $ TeV and the backward region $ -5<\eta^\gamma<-4 $
are shown in Fig.~\ref{fig:fig8} where the black solid, blue dashed, pink dotted-dashed
and green dotted-dotted-dashed curves correspond to the nCTEQ15, EPS09, DSSZ and
HKN07 predictions, respectively, and the free proton PDFs have been taken again from the CT14.
The red band corresponds to the nCTEQ15 NPDF uncertainties. As expected, the EPS09, DSSZ and
HKN07 predictions are in good agreement with each other and nCTEQ15 has significant deviations from them.
It should be noted that only the nCTEQ15 predicts a value greater than 1 for $ R_{p\textrm{Pb}}^\gamma $
in this kinematic region. Moreover, the NPDF error band of nCTEQ15 
is clearly smaller than the corresponding band in the forward direction (see Fig.\ref{fig:fig6}).
This is due to the fact that the NPDFs are constrained better in the antishadowing and EMC effect
regions than the shadowing with experimental data now available. 

\begin{figure}[!]
\centering
\includegraphics[width=8.6cm]{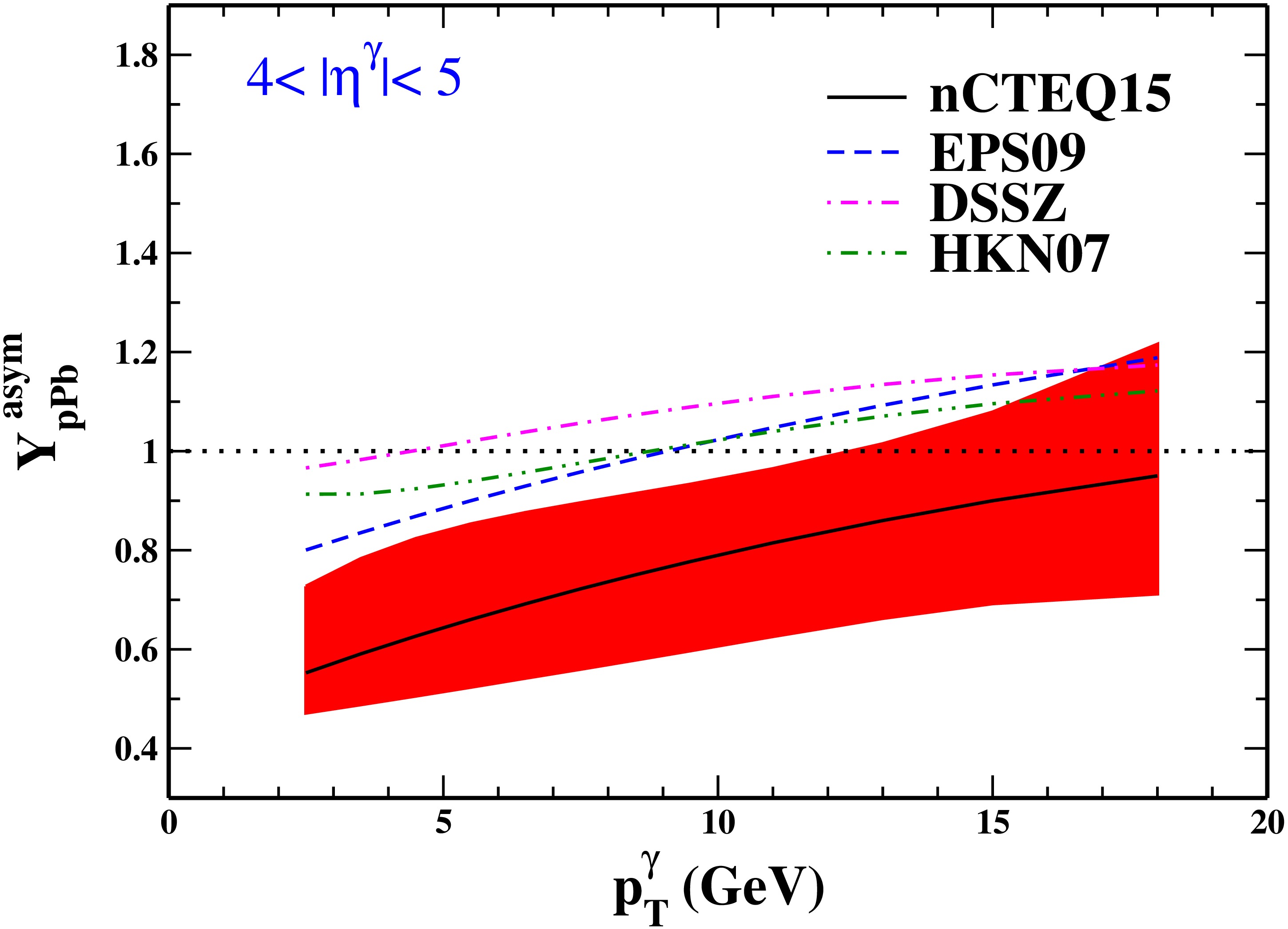}
\caption{A comparison between the forward-to-backward yield asymmetries $ Y_{p\textrm{Pb}}^\textrm{asym} $
for the isolated prompt photon production in $ p $-Pb collisions at $ \sqrt{s}=8.8 $ TeV and $ 4<|\eta^\gamma|<5 $
using the nCTEQ15~\cite{Kovarik:2015cma} (black solid), EPS09~\cite{Eskola:2009uj} (blue dashed), 
DSSZ~\cite{deFlorian:2011fp} (pink dotted-dashed) and HKN07~\cite{Hirai:2007sx} (green dotted-dotted-dashed) 
nuclear modifications and CT14 free-proton PDFs~\cite{Dulat:2015mca}.
The red band corresponds to the nCTEQ15 NPDF uncertainties.}
\label{fig:fig9}
\end{figure}

The corresponding results of the forward-to-backward yield asymmetries $ Y_{p\textrm{Pb}}^\textrm{asym} $
for the isolated prompt photon production in $ p $-Pb collisions 
at $ \sqrt{s}=8.8 $ TeV and $ 4<|\eta^\gamma|<5 $ are shown in Fig.~\ref{fig:fig9}.
The predictions have been made again using the nCTEQ15 (black solid), EPS09 (blue dashed), 
DSSZ (pink dotted-dashed) and HKN07 (green dotted-dotted-dashed) nuclear modifications and the CT14 free-proton PDFs.
In comparison with the nuclear modification ratios (Fig.~\ref{fig:fig6}), 
the nCTEQ15 still has the greatest difference with the others and its prediction is placed below them
in all values of $ p_\textrm{T}^\gamma $. It is interesting that in this case all three
EPS09, DSSZ and HKN07 predictions are not even within the error band of nCTEQ15.
Note also that the nCTEQ15 does not reach 1 even at high values of $ p_\textrm{T}^\gamma $
while the other groups predict a value greater than 1 for $ Y_{p\textrm{Pb}}^\textrm{asym} $
almost at $ p_\textrm{T}^\gamma\gtrsim 9 $ GeV. However, the NPDF error band of nCTEQ15 (red band) 
has not changed significantly except for very low $ p_\textrm{T}^\gamma $ region. 
It is worth remembering that the yield asymmetries are sensitive to two very different $ x_2 $ regions.
Nevertheless, overall, a partial cancellation of the uncertainties in $ Y_{p\textrm{Pb}}^\textrm{asym} $
occurs if the forward and backward nuclear modification ratios are sensitive to the same nuclear effect.
Base on the results obtained in this section, one can simply conclude that
in order to accurately determine the NPDFs and judgements about which one of them 
has more accurate behavior in various $ x $ regions, the measurements of
$ Y_{p\textrm{Pb}}^\textrm{asym} $ are more preferred especially if done with sufficient accuracy. 

\begin{figure}[!]
\centering
\includegraphics[width=8.6cm]{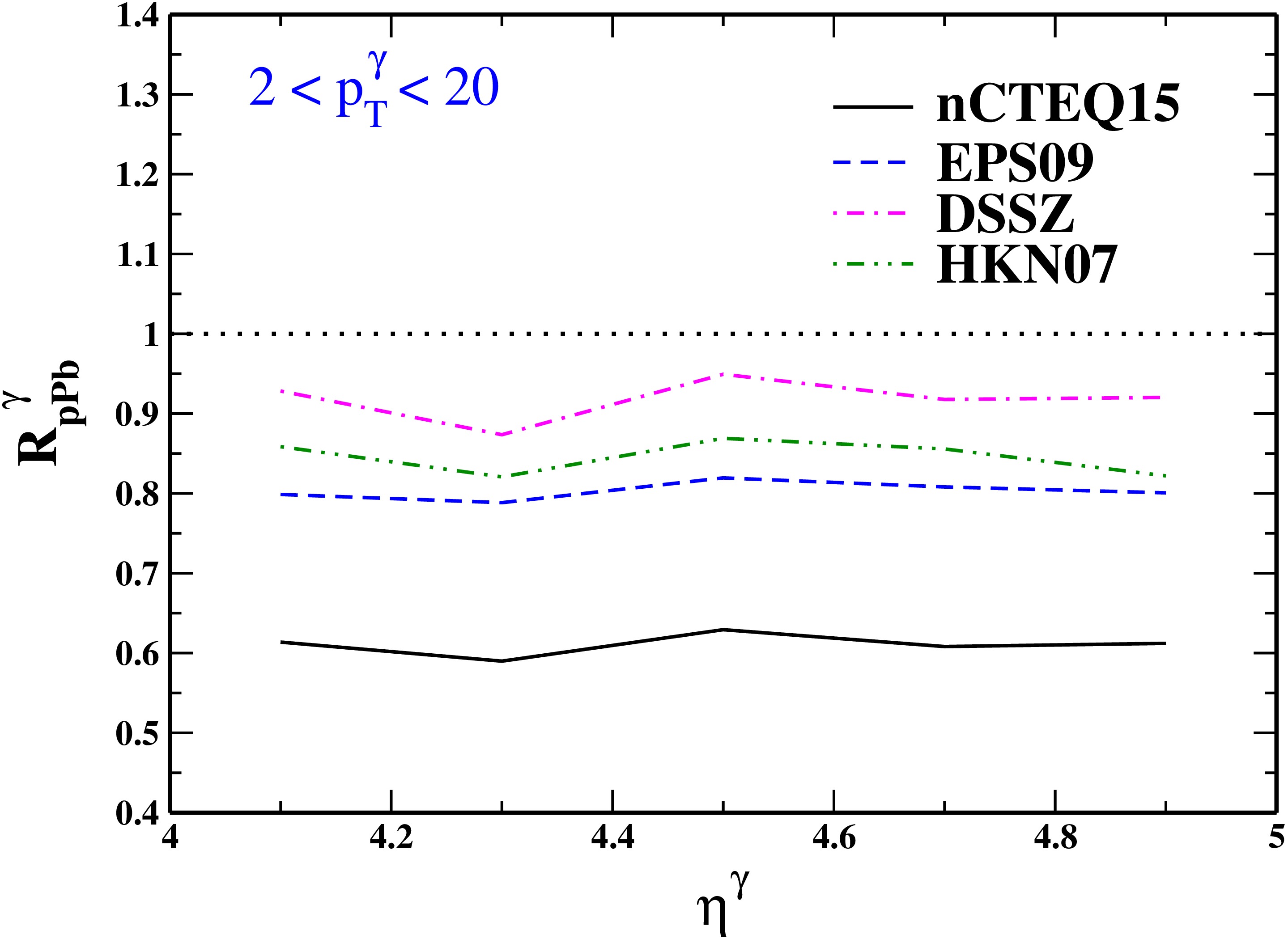}
\caption{A comparison between the nuclear modification ratios $ R_{p\textrm{Pb}}^\gamma $
as a function of $ \eta^\gamma $ for $ p $-Pb collisions at $ \sqrt{s}=8.8 $ TeV with $ 2<p_\textrm{T}^\gamma<20 $ GeV 
using the nCTEQ15~\cite{Kovarik:2015cma} (black solid), EPS09~\cite{Eskola:2009uj} (blue dashed), 
DSSZ~\cite{deFlorian:2011fp} (pink dotted-dashed) and HKN07~\cite{Hirai:2007sx} (green dotted-dotted-dashed) 
nuclear modifications and CT14 free-proton PDFs~\cite{Dulat:2015mca}.}
\label{fig:fig10}
\end{figure}

As a last step, to further explore the impact of input nuclear modifications on the
cross section of isolated prompt photon production in $ p $-Pb collisions, we calculate 
the nuclear modification ratios $ R_{p\textrm{Pb}}^\gamma $ as a function of photon
pseudorapidity $ \eta^\gamma $ using various NPDFs and compare them with each other.
Fig.~\ref{fig:fig10} shows the results obtained at $ \sqrt{s}=8.8 $ TeV for $ 2<p_\textrm{T}^\gamma<20 $ GeV 
and in the kinematic range $ 4<\eta^\gamma<5 $.
The nCTEQ15, EPS09, DSSZ and HKN07 predictions are shown with the 
black solid, blue dashed, pink dotted-dashed
and green dotted-dotted-dashed curves, respectively. As can be seen,
in analogy with Fig.~\ref{fig:fig6}, the differences between the predictions are a little clearer
in this case. Therefore, the measurements of $ R_{p\textrm{Pb}}^\gamma $ as a function of $ \eta^\gamma $
can also be helpful in the determination of nuclear modifications.
%

%
\section{Summary and conclusions}\label{sec:seven}
Photon production in hadron collisions is known as an important tool for testing perturbative QCD predictions.
One of the main motivations of the study of prompt photon production is that 
it is very useful to obtain direct information on the gluon PDFs of both nucleons 
and nuclei. Since the NPDFs cannot be well determined using the available experimental data compared 
with the PDFs of free nucleon, the obtained NPDFs from 
different global analyses by various groups have some considerable differences 
both in behavior and uncertainty. This can lead to the different results for
predictions of physical observables that are sensitive to NPDFs.
In this work, we investigated the impact of various recent NPDFs
on the theoretical predictions of isolated prompt photon production in p-Pb collisions at the LHC for the ALICE kinematics
to estimate the order of magnitude of the difference between their predictions.
We also studied in detail the theoretical uncertainties in the cross sections due to the NPDF, scale and FF uncertainties.
We found that there is no significant difference between the predictions 
obtained using different FFs of BFG sets I and II in the ALICE kinematics. The scale uncertainties
are dominant rather than NPDF uncertainties in all the ranges of $ p_\textrm{T}^\gamma $, and they become 
very large at low $ p_\textrm{T}^\gamma $, if one uses the method 
in which the combination of both incoherent and coherent scale variations is considered.
However, if one considers only the coherent scale variations, a narrow error band is obtained 
in almost all $ p_\textrm{T}^\gamma $ regions, so the scale uncertainties do not even exceed 5\%.
Moreover, we found that there is a remarkable difference between the 
predictions from the nCTEQ15 and other groups in all ranges of $ p_\textrm{T}^\gamma $.
Their differences become more explicit in the calculation of the nuclear modification ratio $ R_{p\textrm{Pb}}^\gamma $
and also in the yield asymmetry between the forward and
backward rapidities $ Y_{p\textrm{Pb}}^\textrm{asym} $ rather than a single differential cross section.
For the forward $ R_{p\textrm{Pb}}^\gamma $, the DSSZ prediction is not even within the large
error band of nCTEQ15. Such a situation occurs for the backward $ R_{p\textrm{Pb}}^\gamma $
and also the yield asymmetries $ Y_{p\textrm{Pb}}^\textrm{asym} $ in almost all $ p_\textrm{T}^\gamma $ regions but this time
for all three EPS09, DSSZ and HKN07 predictions. Overall, the NPDF error band of nCTEQ15 
in the backward direction is smaller than the corresponding band in the forward direction.
This is due to the fact that the isolated photon production at backward rapidities is sensitive to the
nuclear antishadowing and EMC effect of NPDFs which are constrained better than the shadowing with the experimental data now available.
However, the NPDF error band of nCTEQ15 from $ R_{p\textrm{Pb}}^\gamma $ to $ Y_{p\textrm{Pb}}^\textrm{asym} $ 
has not changed significantly except for the very low $ p_\textrm{T}^\gamma $ region.
Base on the results obtained, we concluded that
in order to accurately determine the NPDFs and judgements about which one of them 
has more accurate behavior in various $ x $ regions, the measurements of
$ Y_{p\textrm{Pb}}^\textrm{asym} $ are more preferred especially if done with sufficient accuracy.
In addition, the future measurements with ALICE will be very
useful not only for decreasing the uncertainty of the gluon nuclear modification, but also to accurately determine
its central values especially in the shadowing region.
As further investigation, we calculated the nuclear modification ratio $ R_{p\textrm{Pb}}^\gamma $ as a function of photon
pseudorapidity $ \eta^\gamma $ using various NPDFs. We found that the differences between the predictions are a little clearer
in this case. It seems that the measurements of $ R_{p\textrm{Pb}}^\gamma $ as a function of $ \eta^\gamma $
can also be helpful in the determination of nuclear modifications.

%
\acknowledgments
We thank Hamzeh Khanpour and Matthew D. Schwartz for
useful discussions and comments. This project was
financially supported by Semnan University.
%

\end{document}